\documentclass[12pt]{iopart}

\newcommand{\pp}{$p+p$}
\newcommand{\dAu}{$\rm d+Au$}
\newcommand{\AuAu}{$\rm Au+Au$}
\newcommand{\pT}{$p_{T}$}
\newcommand{\meanpt}{\mbox{$\langle p_T \rangle$}}

\newcommand{\sigmastar}{\mbox{$\Sigma  (1385)$}}
\newcommand{\sigmastarbar}{\mbox{$\overline{\Sigma } (1385)$}}

\newcommand{\sqrtsNNB}{$\sqrt{s_{_{\mathrm{NN}}}}$=200 GeV}
\usepackage{graphicx}
\begin{document}

\title[Baryonic Resonance Studies with STAR]{Baryonic Resonance Studies with STAR}

\author{Sevil Salur (for the STAR Collaboration)}

\address{Yale University, 272 Whitney Ave., New Haven, CT 06520 USA} \ead{sevil.salur@yale.edu}
\begin{abstract}
Yields and spectra of \sigmastar\ are measured in \pp, \dAu\ and
\AuAu\ collisions at \sqrtsNNB. The nuclear modification factors
in \dAu\ collisions are presented. The \pT\ dependent medium
effects are investigated via the nuclear modification factors. The
implications of these results on various models are discussed.
\end{abstract}


\section{Introduction}

Strongly interacting, high density matter is produced in heavy ion
collisions at the Relativistic Heavy Ion Collider. Hadronic
resonances, due to their short lifetimes, can be used to
investigate the freeze-out mechanisms after hadronization. The
production of the  strange baryonic resonance \sigmastar\
 is investigated for the first time in heavy ion
collisions and, through comparison with other resonances, the
evolution of the fireball is investigated.  The collision dynamics
are studied with the $\Sigma(1385)/\Lambda$ ratio in comparison to
other resonance/stable particle ratios to explore the
re-scattering and regeneration effects between chemical and
thermal freeze-out \cite{prl,salur}.

\section{Analysis and Particle Identification Techniques}

\begin{figure}[ht]
\centering$\begin{array}{cc}
\includegraphics[height=5.8cm,width=7cm,clip]{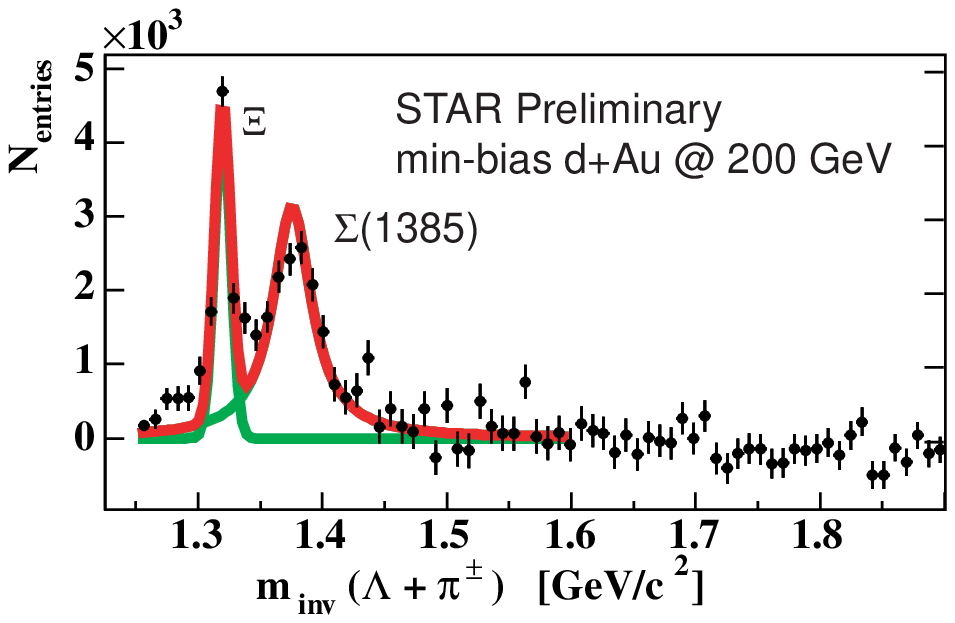}\hspace{-0.1in}
&\includegraphics[height=5.8cm,clip]{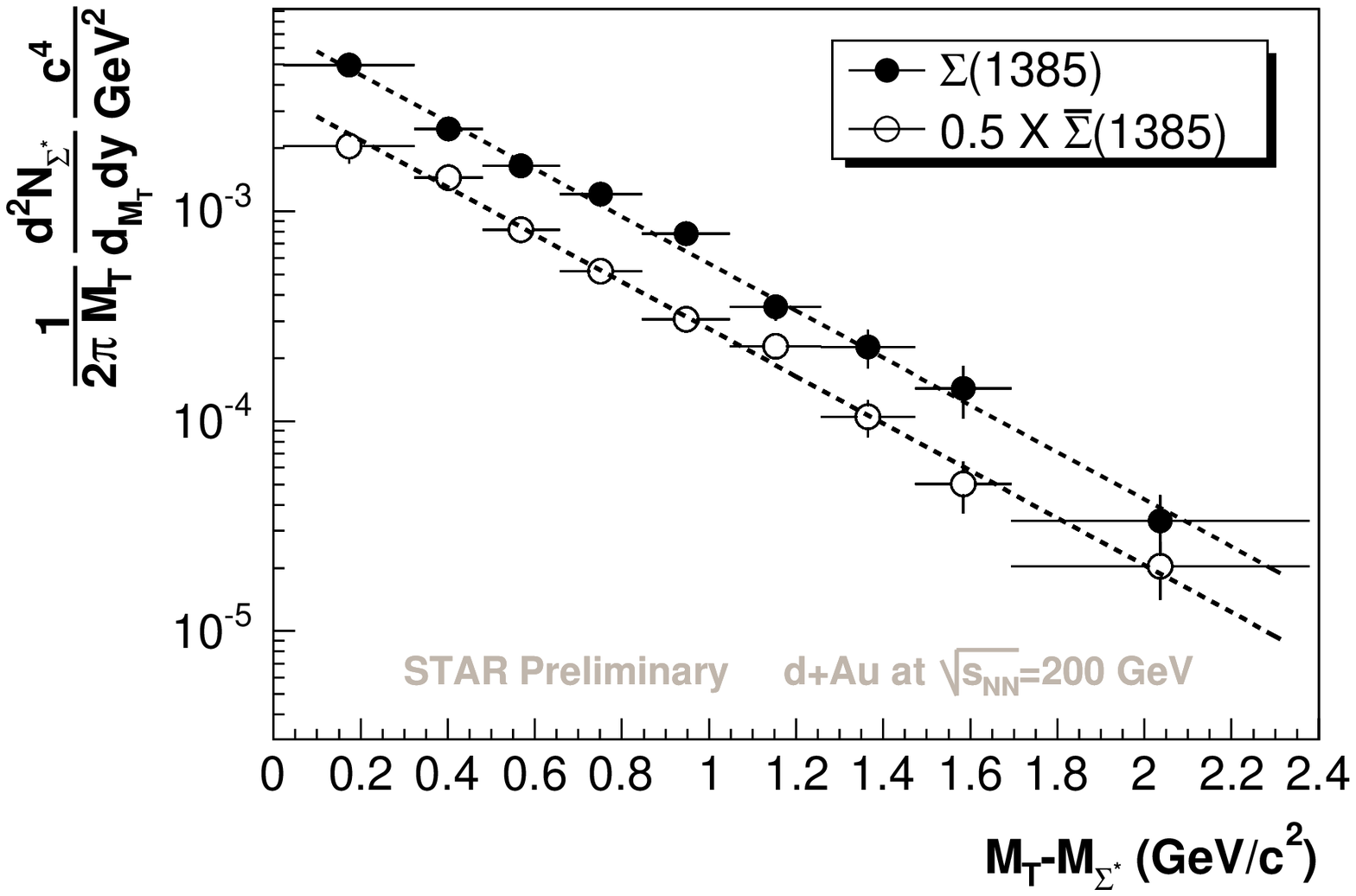} \\
\mbox{\bf   (a) }&\mbox{\bf   (b)}
\end{array}$
 \caption[]{{\bf   (a)} Invariant mass spectra of the \sigmastar\ after
 mixed-event background subtraction in minimum bias \dAu\ collisions. A
Gaussian for $\Xi$, Breit-Wigner for the \sigmastar\ are fit to
the spectra.{\bf   (b) } The transverse mass spectra for
\sigmastar\ in minimum bias \dAu\ collisions at \sqrtsNNB. The
dashed curves represent the exponential fits to the data.}
 \label{fig:dAumtspectra}      
\end{figure}

The direct identification and measurement of the \sigmastar\
($\rightarrow \Lambda +\pi$) in the detectors is not possible due
to its short life-time ($c\tau_{\Sigma (1385) }=6 \;\rm fm$).
Instead, the \sigmastar\ is identified by reconstructing the
invariant mass distribution from its decay products via a
combinatorial technique. In this technique, \sigmastar\ baryons
are identified by combining the topologically reconstructed
$\Lambda$ baryons with $\pi$ mesons that are identified via their
$dE/dx$ and momentum information from the STAR Time Projection
Chamber.
Figure~\ref{fig:dAumtspectra}-a shows the clear signal of the
invariant mass spectrum of \sigmastar\ after mixed-event
background subtraction. Taking into account our detector
resolution and the measurement uncertainties, the measured widths
of the $\Xi^{-}$ and \sigmastar\  are in agreement with the PDG
\cite{pdg}. Corrected mid-rapidity ($|y|<0.5$) $\rm m_{T}- m_{0}$
spectra are presented in Figure~\ref{fig:dAumtspectra}-b for the
\sigmastar\ (closed circles) and \sigmastarbar\ (open circles)
from minimum bias \dAu\ collisions at \sqrtsNNB. The error bars
presented correspond to both statistical and bin-by-bin systematic
uncertainties.

\section{Results and Discussions}

The \pT\ spectra  and \meanpt\ are also measured in \pp\ and
\AuAu\ collision environments \cite{prl,salur}. Table~\ref{table2}
summarizes the results from the corrected spectra of the
\sigmastar\ with the statistical and systematic errors in all
three collision systems.

\begin{table*}[h!]
 \centering
 \caption{ The $\langle
p_{T}\rangle $ and yield ($dN/dy$) from exponential fits to the
\sigmastar\ ($\Sigma ^{*}$) $p_{T}$ spectra ~\cite{prl,salur}. The yields for \pp\ are from non-singly diffractive collisions. }  
 \label{table2}
\begin{tabular}{lccccc}
Particle & Collision & Centrality &  $\langle p_{T}$$\rangle$ [GeV/c] & $(dN/dy)|_{y=0}$ \\
  \hline
$\Sigma ^{* \pm}$ &  $p+p$ & min-bias &  $1.02 \pm 0.02 \pm 0.07$  & $(10.7 \pm 0.4 \pm 1.4)\times10^{-3}$  \\
$\overline \Sigma ^{* \pm}$ &  $p+p$  & min-bias  & $1.01 \pm 0.01 \pm 0.06$  & $(8.9 \pm 0.4 \pm 1.2)\times10^{-3}$  \\
$\overline{\Sigma}^{* \pm}+\Sigma^{* \pm}$ &  Au+Au &  0-5\% &  $1.28 \pm 0.15 \pm 0.09$ & $9.3 \pm 1.4 \pm 1.2$ \\
$\Sigma ^{* \pm}$ &  $\rm d+Au$ & min-bias &  $1.14 \pm 0.05 \pm 0.08$  & $(3.23 \pm 0.15 \pm 0.42)\times10^{-2}$  \\
$\overline \Sigma ^{* \pm}$ &  $\rm d+Au$  & min-bias  & $1.12 \pm 0.05 \pm 0.08$  & $(3.15 \pm 0.15 \pm 0.41)\times10^{-2}$  \\
\end{tabular}
\end{table*}

\begin{figure}[h!]
\centering$\begin{array}{cc}
\includegraphics[height=6.3cm,clip]{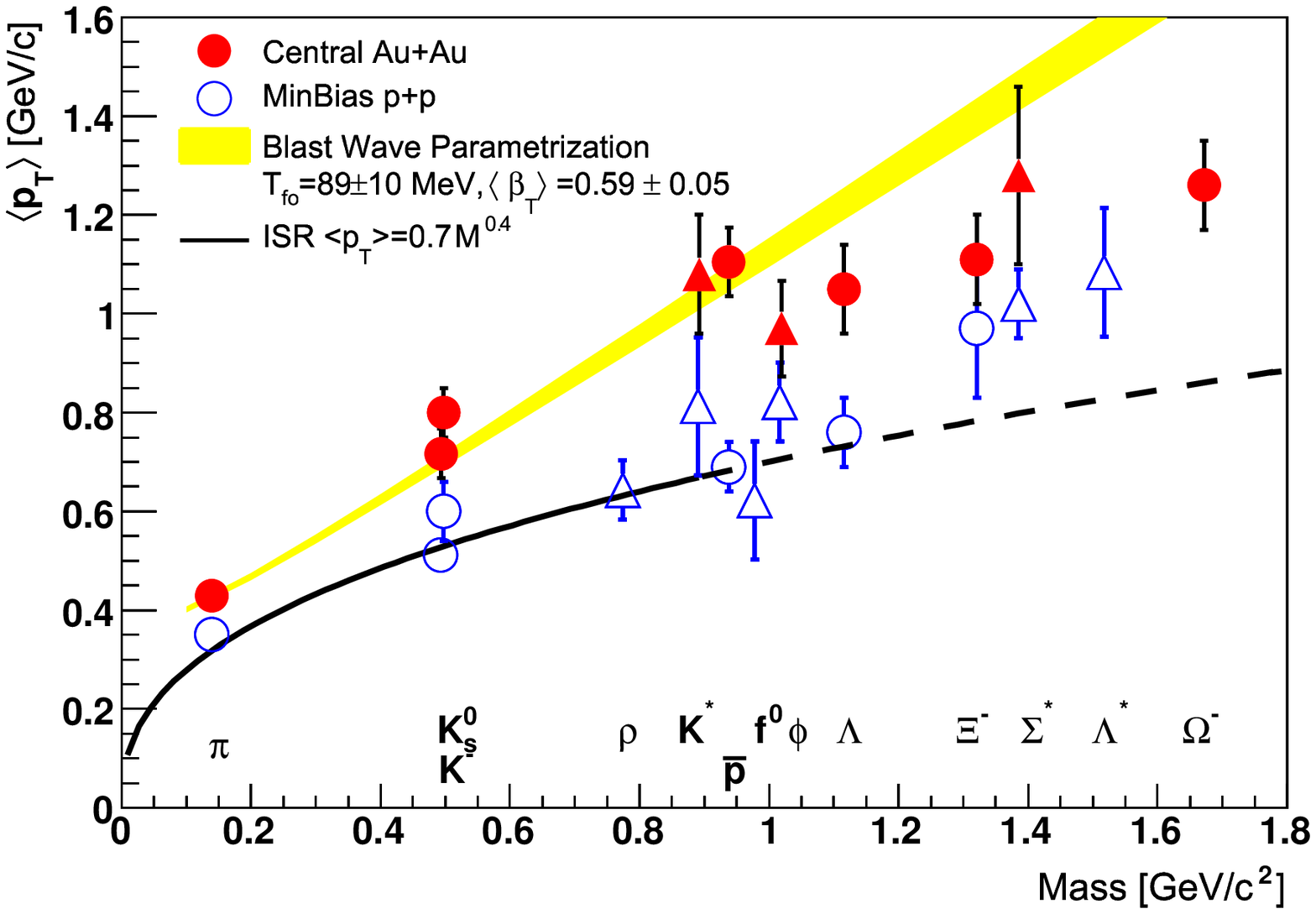}\hspace{-0.2in}
&\includegraphics[height=6.3cm,clip]{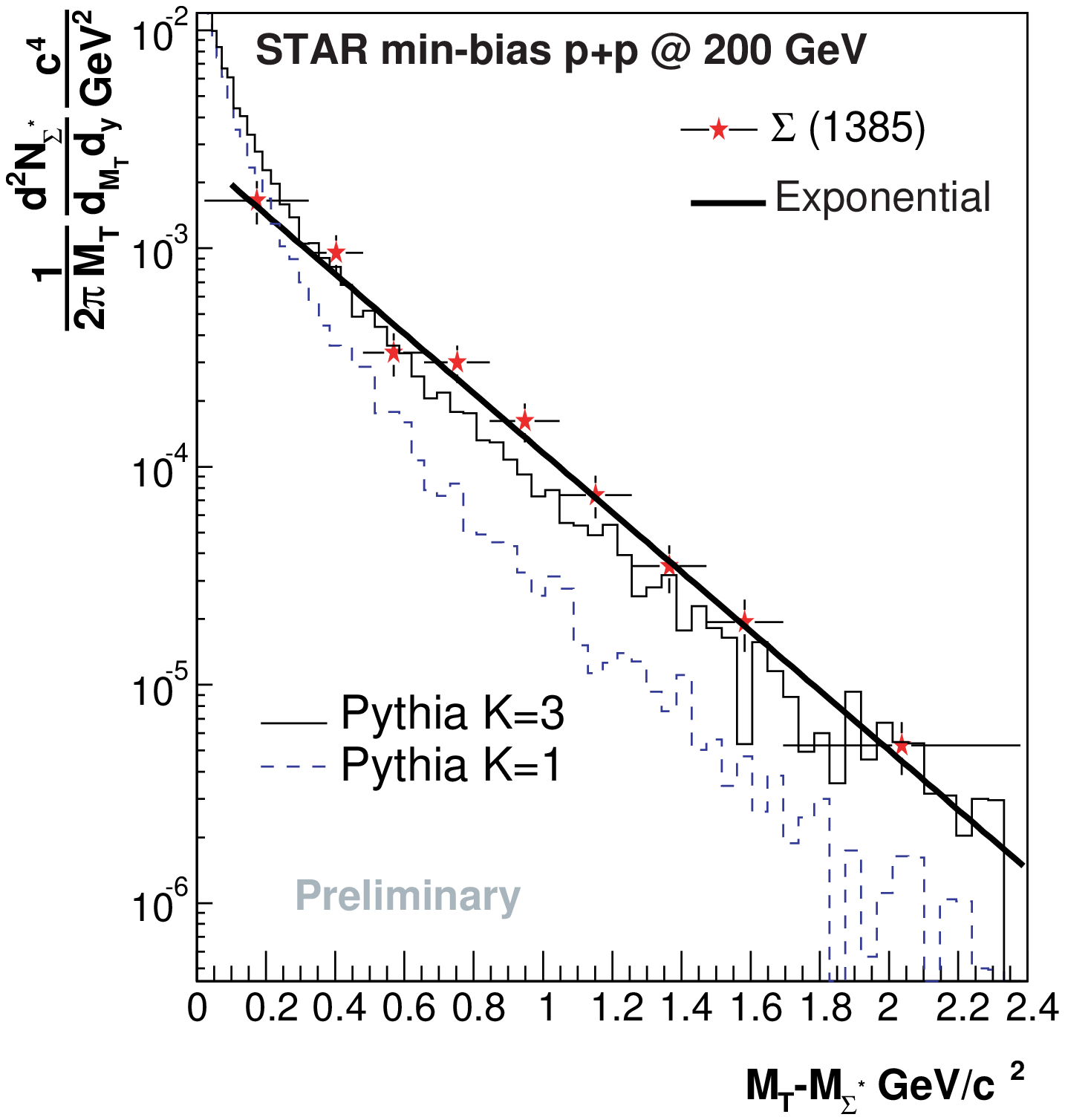} \\
\mbox{\bf   (a) }&\mbox{\bf   (b)}
\end{array}$

\caption[]{{\bf   (a) }The $\langle p_{T} \rangle$  vs particle
mass measured in \pp\ and \AuAu\ collisions at \sqrtsNNB.  {\bf
(b) } Comparison of the \sigmastar\ spectrum with Pythia
predictions for two different values of the K factor. See text for
details.}
\label{massvspt}       
\end{figure}

A comparison of the \meanpt\ as a function of measured particle
mass is presented in Figure~\ref{massvspt}-a for \pp\ and \AuAu\
collisions at \sqrtsNNB.   The triangles represent the resonances
and the circles signify long-lived `stable' particles. The black
curve is an empirical fit to the ISR $\pi$, K, and p data in \pp\
collisions and the band is a blastwave fit using $\pi$, K, and $p$
in STAR for Au+Au collisions \cite{ISR01,mult130}. The empirical
parametrization for the ISR data at $\sqrt{s}=$ 25 GeV in \pp\
collisions, can describe the \meanpt\ of the lower mass particles,
such as $\pi$, K, and $p$, despite the fact that our collision
energy is an order of magnitude higher. However, this empirical
parametrization does not reproduce the behavior of the higher mass
particles in \pp\ collisions. Similarly, the blastwave
parametrization which can describe the \meanpt\ of the lower mass
particles ($\sim 98\%$ of all the particles observed) in \AuAu\
collisions, fails for the higher mass particles.

The heavy particles in \pp\ and \AuAu\ collisions show a similar
magnitude of $\langle p_{T} \rangle$. It is expected that
resonances with higher transverse momentum are more likely to be
reconstructed because of their longer relative lifetimes due to
their ultra-relativistic velocities. This means they are more
likely to decay outside  the medium and hence their daughter
particles will interact less with the medium in \AuAu\ collisions.
Any loss of low $p_{T}$ \sigmastar\ baryons will cause an increase
in the measured \meanpt\ of the observed $p_{T}$ spectra for the
central \AuAu\ collisions with respect to \pp\ collisions. However
we do not see any significant increase in the \meanpt\ for the
$\Sigma (1385)$ from minimum bias \pp\ to the most central \AuAu\
collisions within the statistical and systematic errors. It is
possible that the production of the higher mass particles in \pp\
collisions is biased towards higher multiplicity collisions. If
the higher mass particles are produced in more violent (mini-jet)
\pp\ collisions compared to lower mass particles, the $\langle
p_{T} \rangle$ for heavy particles in \pp\ collisions would be
larger.

A comparison of the \sigmastar\ spectrum with a leading order pQCD
model, Pythia 6.3~\cite{ref:pythia}, is presented in
Figure~\ref{massvspt}-b. It is possible to model the \sigmastar\
spectrum with Pythia with the factor K=3 while the default (K=1)
spectrum is too soft. This K-factor, which represents a simple
factorization of next-to-leading order processes (NLO) in the
Pythia leading order (LO) calculation, is directly related to the
\pT\ spectrum. An increase in the K factor implies that larger NLO
contributions (mini-jet events) are required to describe the
\sigmastar\ production. The high (K=3) factor is also needed to
describe heavy strange baryons such as the $\Lambda$ and the
$\Xi$, while the spectra predicted with this high K factor are too
hard for the light mesons \cite{ref:mark}. In \AuAu\ collisions,
there is evidence that heavier particles flow radially with a
smaller boost velocity due to their smaller cross section than the
lighter mass particles. These two independent effects in \pp\ and
\AuAu\ collisions may cause the apparent merging of the $\langle
p_{T} \rangle$.

 In a thermal model,
the measured ratios of resonance to non-resonant particles with
identical valence quarks are sensitive to the chemical freeze-out
temperature, as all of the quark content dependencies cancel out
\cite{tor01}. While these models predict the measured
$\Sigma(1385)/\Lambda$ ratio correctly within errors, they suggest
a higher ratio than the measured $\Lambda (1520)/\Lambda$ in the
most central \AuAu\ collisions. This suggests an extended hadronic
phase of elastic and pseudo-elastic interactions after chemical
freeze-out, where re-scattering of resonance decay particles and
regeneration of resonances  will occur.

\begin{figure}[h!]
\centering
\resizebox{0.96\textwidth}{!}{%
  \includegraphics{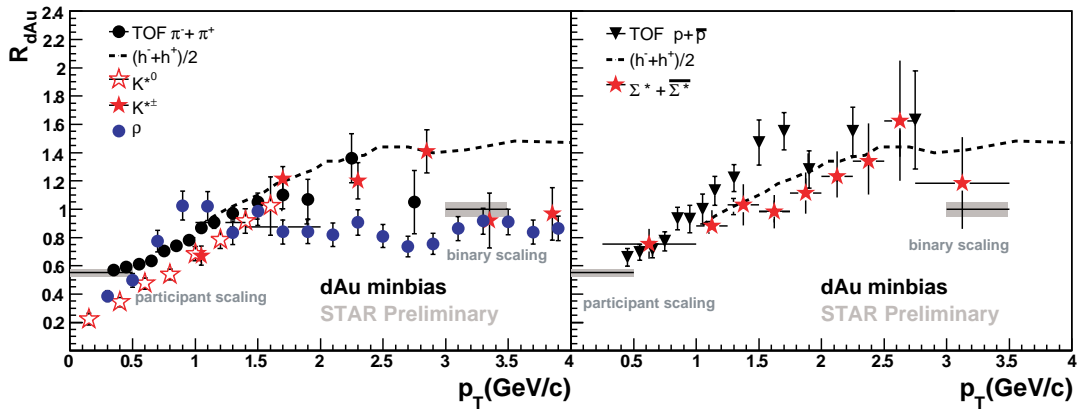}
}\caption[]{Nuclear modification factors ($R_{dAu}$) for the
\sigmastar\ in comparison to other mesons on the left and baryons
on the right in \dAu\ collisions.}
\label{fig:nuclearmoddauall}       
\end{figure}

The nuclear modification factor for the \sigmastar\ in comparison
to other mesons and baryons in \dAu\ collisions can be found in
Figure~\ref{fig:nuclearmoddauall}. The $R_{dAu}$ measurements, for
mesons on the left and for baryons on the right, mostly follow
participant scaling at low momenta. At higher momentum, baryons
show a greater enhancement over the binary scaling than the
mesons. The \sigmastar\ baryon follows a similar trend to $\rm
h^{\pm}$. The enhancement over binary scaling can be described by
the Cronin effect, a generic term for the experimentally observed
broadening of transverse momentum spectra at intermediate \pT\ in
$p$+A collisions as compared to \pp. It is surprising that the
$\rho$ meson shows no enhancement above binary collisions and even
falls below $\pi$ mesons, while the other resonances and their
stable particles show no clear difference in their $R_{dAu}$.

\end{document}